\documentclass[
    ,final            
  ]
  {aipproc}

\layoutstyle{8x11single}

\begin{document}

\title{A Search for Gamma-ray Burst Subgroups in the SWIFT and RHESSI Databases}


\author{Jakub \v{R}\'{i}pa}{address={Astronomical Institute of the Charles University, V Hole\v{s}ovi\v{c}k\'{a}ch 2, Prague, Czech Republic}}

\author{David Huja}{address={Astronomical Institute of the Charles University, V Hole\v{s}ovi\v{c}k\'{a}ch 2, Prague, Czech Republic}}

\author{Attila M\'{e}sz\'{a}ros}{address={Astronomical Institute of the Charles University, V Hole\v{s}ovi\v{c}k\'{a}ch 2, Prague, Czech Republic}}

\author{Ren\'{e} Hudec}{address={Astronomical Institute, Academy of Sciences of the Czech Republic, Ond\v{r}ejov, Czech Republic}}

\author{Wojtek Hajdas}{address={Paul Scherrer Institut, Villigen, Switzerland}}

\author{Claudia Wigger}{address={Paul Scherrer Institut, Villigen, Switzerland},altaddress={Kantonsschule Wohlen, Switzerland}}

\begin{abstract}
 A sample of 286 gamma-ray bursts (GRBs) detected by the Swift satellite and 358 GRBs detected by the RHESSI satellite are studied statistically. Previously published articles, based on the BATSE GRB Catalog, claimed the existence of an intermediate subgroup of GRBs with respect to duration. We use the statistical $\chi^2$ test and the F-test to compare the number of GRB subgroups in our databases with the earlier BATSE results. Similarly to the BATSE database, the short and long subgroups are well detected in the Swift and RHESSI data. However, contrary to the BATSE data, we have not found a statistically significant intermediate subgroup in either Swift or RHESSI data.
\end{abstract}

\maketitle

\section{Introduction}

    By studying the duration distribution of Gamma-Ray Bursts (GRBs), it was originally found (results from BATSE and Konus-Wind instruments \cite{1,2}) that there exist two subclasses; the short one with typical durations of less than 2\,s and the long one with GRB lasting more than 2\,s. However, some articles point to existence of three subclasses of GRBs in the BATSE database with respect to their durations \cite{3,4}.
    Works \cite{9} and \cite{10} using a multivariate analysis reveals at least three groups of GRBs in the BATSE data.
    The article \cite{6} says that the third subclass (with intermediate duration), observed by BATSE, is a bias caused by an instrumental effect.
    Therefore, we decided to investigate durations and number of groups of GRBs in the RHESSI and Swift databases.

\section{RHESSI and Swift Instruments}

   The Ramaty High Energy Solar Spectroscopic Imager (RHESSI) is a NASA Small Explorer satellite designed to study hard X-rays and gamma-rays from solar flares (see at: \url{http://hesperia.gsfc.nasa.gov/hessi}). It consists mainly of an imaging tube and a spectrometer. The spectrometer consists of nine germanium detectors (7.1\,cm diameter and 8.5\,cm height). They are lightly shielded only, thus making RHESSI also very useful to detect non solar photons from any direction. The energy range for GRB detection extends from about 50\,keV up to 17\,MeV depending on the direction. The energy and time resolutions are: $\Delta E$~=~3\,keV (at 1000\,keV), $\Delta t$~=~1\,$\mu$s. An effective area reaches up to 150\,cm$^2$ at 200\,keV. With a field of view of about half of the sky, RHESSI observes about one or two gamma-ray bursts per week.

   The Swift satellite is described at \url{http://swift.gsfc.nasa.gov/docs/swift}.

\section{Data Samples}

   Two samples are used. The first is the set of 358 GRBs observed by the RHESSI satellite and covers the period from February 2002 to April 2008 (see: \url{http://grb.web.psi.ch}). We have used the SSW program under IDL and authors' routines to derive count light-curves in the energy range  25 - 1500\,keV.

   The second data-sample is the set of 286 GRBs detected by the Swift satellite. The sample covers the period from November 2004 to December 2007 (see at: \url{http://swift.gsfc.nasa.gov/docs/swift/archive/grb_table}). The Swift energy range is 15 - 150 keV.

   Here we present a comparison of the statistical analysis of these two samples and discuss our results with similar analyses of the BATSE Catalog done by others (\cite{3}-\cite{5}).

\section{Duration Distribution}

\begin{figure}
  \includegraphics[width=\textwidth]{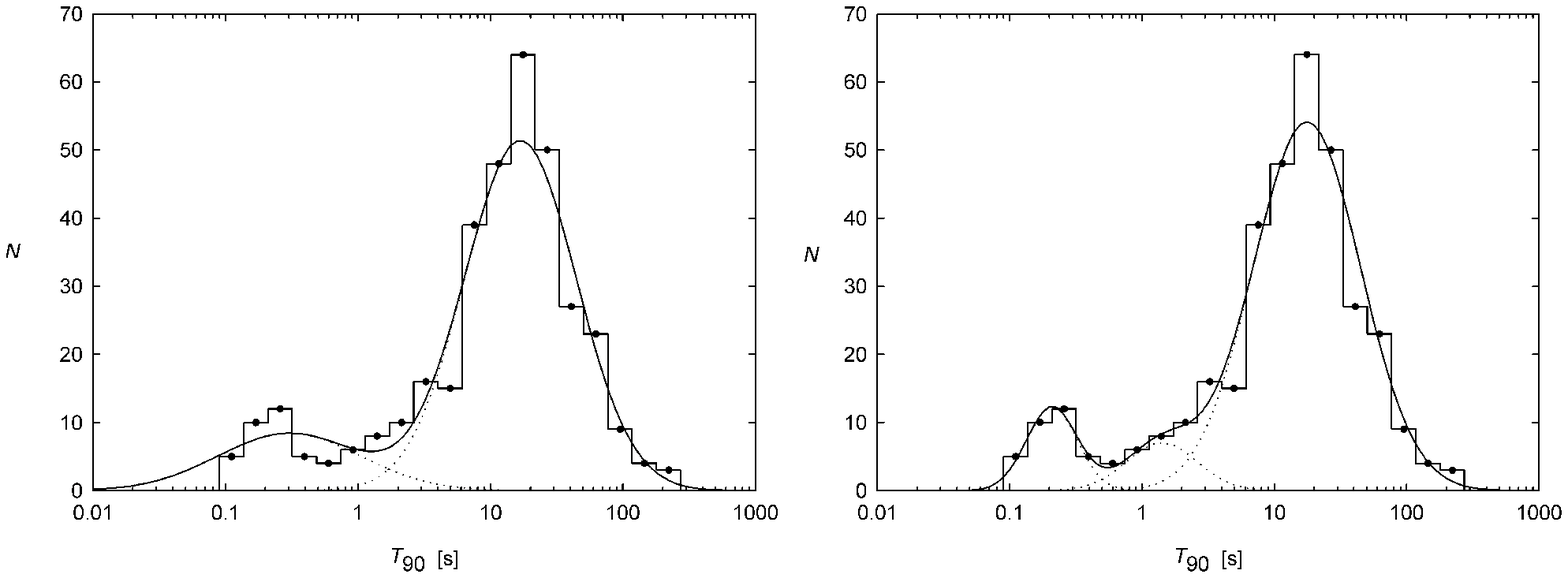}
  \caption{Distribution of the GRB RHESSI durations with 2 log-normal (left) and 3 log-normal (right) fits.}
  \label{T90-RHESSI}
\end{figure}

\begin{figure}
  \includegraphics[height=0.25\textheight]{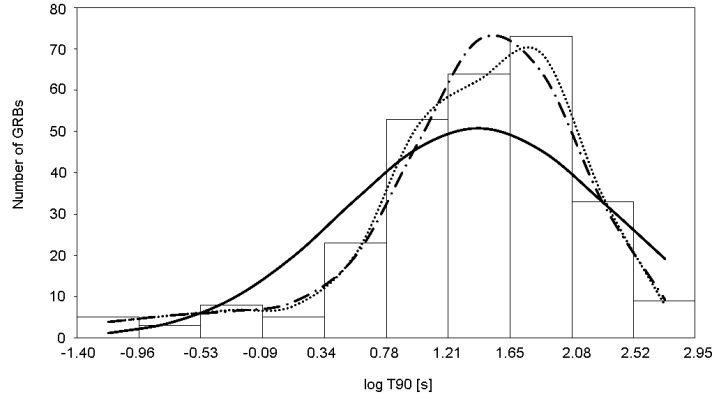}
  \caption{Distribution of the GRB Swift durations with 1 log-normal (full line), 2 log-normal (dot-dashed line)
           and 3 log-normal (dotted line) fits.}
  \label{T90-Swift}
\end{figure}

    The duration distributions of GRBs observed by RHESSI and SWIFT are
    shown in Figure~\ref{T90-RHESSI} and Figure~\ref{T90-Swift}, respectively.
    As duration, we use $T_{90}$ i.e. the time
    interval during which the cumulative counts increase
    from 5\% to 95\% above a background.

    We followed the method done in \cite{3} and fitted the data by one, two and three log-normal functions.
    We used the $\chi^2$ test to evaluate these fits, the measurement errors being the square root of the number
    of GRBs per bin.

    In the case of one log-normal fit for both samples we obtained the goodness of fit < 0.01 \%. Therefore, the assumption that there is only one subclass can be rejected.
   
    The parameters of the other fits, together with
    the $\chi^2$ values and goodness-of-fits, are
    noted in Table~\ref{tabulka}.

    The question is whether the improvement in $\chi^2$, when adding a third log-normal function, is statistically significant.
    To answer this question, we used the F-test as described in \cite{7}.
    In the case of the RHESSI data, the obtained critical value $F_{0}$ implies the probability $P$($F$>$F_{0}$) = 8.6 \% that
    the improvement is just a statistical fluctuation. This probability is still too high to doubtlessly reject the hypothesis
    that the improvement in $\chi^2$ is being accidentally.

    The situation with the Swift data-set is: the fit with three log-normal functions gives a worse
    goodness-of-fit than the one with two log-normal functions.
    Hence, a third subgroup is not found.

\section{Discussion and Conclusion}

Our result does not confirm
results in the work \cite{3} (done with BATSE data),
where the goodnesses of the two and
three log-normal fits  gave values 40 \% and 98 \%,
respectively, and statistically proved the hypothesis of
the third intermediate GRB subclass.

We can not doubtlessly confirm the intermediate subclass of GRBs
in the RHESSI sample by using $\chi^2$ fitting and F-test.
Also in the case of the Swift data-sample, we can not confirm the intermediate subclass of GRBs.
That is in the contrary to a recent work \cite{8}.
That work, using the maximum likelihood ratio test, reveals statistically significant intermediate GRB group
in the Swift data-set.
We are aware of the low counts in the histogram bins (because of our data-sets sizes)
and this fact can negatively affect our results.

Another result is the number of GRBs with $T_{90}$ < 2 s
which is for RHESSI $\simeq 14.0$ \%, for Swift $\simeq 7.6$ \%
and for BATSE 4B $\simeq 26$ \%;
therefore, the ratio short/long GRBs obviously depends on the instrument.

\begin{center}
\begin{table} \caption{Parameters of 2 and 3 log-normal fits of $T_{90}$ of the RHESSI and Swift
                       data-sets. $\mu$ are the means, $\sigma$ are standard deviations, $w$ are
                       the weights of the distribution and DoF means degree-of-freedom.}
\centering
\begin{tabular}{lrrrr}

\hline
\\[-2ex]
parameter                    &  RHESSI 2-fit    &   RHESSI 3-fit   &  Swift 2-fit    &   Swift 3-fit \\[0.3ex]
\hline\hline
\\[-2ex]
$\mu_{\textrm{short}}$       &         -0.52    &            -0.68 &      -0.27      &      0.31     \\[0.3ex]
$\sigma_{\textrm{short}}$    &          0.54    &             0.19 &       0.93      &      1.27     \\[0.3ex]
$w_{\textrm{short}}[\%]$     &          17.3    &            9.0   &       12.2      &      20.2     \\[0.3ex]
\hline
$\mu_{\textrm{long}}$        &          1.23    &             1.25 &       1.57      &      1.72     \\[0.3ex]
$\sigma_{\textrm{long}}$     &          0.42    &             0.41 &       0.56      &      0.46     \\[0.3ex]
$w_{\textrm{long}}[\%]$      &          82.7    &            83.8  &       87.8      &      67.9     \\[0.3ex]
\hline
$\mu_{\textrm{middle}}$      &                  &             0.15 &                 &      0.82     \\[0.3ex]
$\sigma_{\textrm{middle}}$   &                  &             0.27 &                 &      0.08     \\[0.3ex]
$w_{\textrm{middle}}[\%]$    &                  &            7.2   &                 &      11.9     \\[0.3ex]
\hline
$dof$                        &          14      &            11    &        5        &       2       \\[0.3ex]
$\chi^2$                     &          17.52   &            9.86  &        6.72     &       3.82    \\[0.3ex]
goodness[\%]                 &          22.9    &            54.3  &        24.2     &       14.8    \\[0.3ex]
\hline
\\[-2ex]
$ F_{0} $                    &          2.85    &                  &                 &               \\[0.3ex]
$P(F>F_{0})[\%] $            &           8.6    &                  &                 &               \\[0.3ex]
\\[-2ex]
\hline
\label{tabulka}
\end{tabular}
\end{table}
\end{center}

\begin{theacknowledgments}
This study was supported by the GAUK grant No. 46307, by the OTKA grant No. T48870, by the Grant Agency of the Czech Republic, grant No. 205/08/H005, and by the Research Program MSM0021620860 of the Ministry of Education of the Czech Republic.
\end{theacknowledgments}

\end{document}